\RequirePackage{snapshot}
\pdfoutput=1
\documentclass[a4paper,11pt]{article}

\usepackage{url}

\usepackage{fullpage}

\usepackage{amsmath,amsthm,amssymb}
\usepackage{xcolor}
\usepackage{graphicx}
\usepackage{bbold}
\usepackage{textcomp} 
\usepackage{siunitx}

\usepackage[position=top]{subfig}
\usepackage{authblk}
\renewcommand\Affilfont{\itshape\small}

\usepackage{todonotes}
\makeatletter
\renewcommand{\todo}[2][]{\tikzexternaldisable\@todo[#1]{#2}\tikzexternalenable}
\renewcommand{\missingfigure}[2][]{\tikzexternaldisable\@missingfigure[#1]{#2}\tikzexternalenable}
\makeatother

\title{A three-dimensional spin-diffusion model for micromagnetics}

\author[1]{Claas Abert\thanks{claas.abert@tuwien.ac.at}}
\author[2]{Michele Ruggeri}
\author[1]{Florian Bruckner}
\author[3]{Christoph Vogler}
\author[4]{Gino Hrkac}
\author[2]{Dirk Praetorius}
\author[1]{Dieter Suess}
\affil[1]{Christian Doppler Laboratory of Advanced Magnetic Sensing and Materials, Institute of Solid State Physics, Vienna University of Technology, Austria}
\affil[2]{Institute for Analysis and Scientific Computing, Vienna University of Technology, Austria}
\affil[3]{Institute of Solid State Physics, Vienna University of Technology, Austria}
\affil[4]{College of Engineering, Mathematics and Physical Sciences, University of Exeter, United Kingdom}

\begin{document}
\maketitle

\begin{abstract}
We solve a time-dependent three-dimensional spin-diffusion model coupled to the Landau-Lifshitz-Gilbert equation numerically. The presented model is validated by comparison to two established spin-torque models: The model of Slonzewski that describes spin-torque in multi-layer structures in the presence of a fixed layer and the model of Zhang and Li that describes current driven domain-wall motion. It is shown that both models are incorporated by the spin-diffusion description, i.e., the nonlocal effects of the Slonzewski model are captured as well as the spin-accumulation due to magnetization gradients as described by the model of Zhang and Li. Moreover, the presented method is able to resolve the time dependency of the spin-accumulation.

\end{abstract}

\newpage
\section*{Introduction}
An electric or spin current of spin angular momentum acting on a ferromagnet exerts a torque on the magnetization, by doing that driving it out of equilibrium \cite{slonczewski1996current,berger1996emission}.
Due to its non-conservative nature this spin torque can act either as effective negative or positive magnetic damping and thereby excite magnetization self-oscillations \cite{kiselev2003microwave,slavin2009nonlinear}.
Spin-torque oscillators (STO) have been realized in nanoscale spin valve systems \cite{ozyilmaz2004current,mistral2006current,braganca2010nanoscale} ranging from point contact magnetic multilayer structures \cite{rippard2004direct,mohseni2013spin} to nanoscale magnetic tunnel junction systems \cite{nazarov2006spin,deac2008bias,rowlands2013time}.
Recently, a new type of nano spin-torque oscillator based on current-induced spin orbit torques in a permalloy(Py)/platinum(Pt) bilayer systems was introduced \cite{demidov2012magnetic,liu2013spectral}.
Such spin orbit torques originate from the spin Hall effect in Pt \cite{dyakonov1971possibility,hirsch1999spin,hoffmann2013spin} and the Rashba effect at the Pt/Py interface \cite{obata2008current,miron2010current}.
Applications for STOs include various nanoscale communication technologies.
In the upcoming field of microwave assisted recording, a STO is used to generate fields in the GHz regime in order to locally assist the reversal process in hard disks \cite{zhu2008microwave}.
Another important application of spin torque is to switch magnetic layers in nano structures \cite{hosomi2005novel}.

In order to obtain deeper understanding of the applications utilizing spin torque and to guide the design of novel devices, various numerical models have been developed:
A model for the description of magnetic multi-layers was first introduced by Slonczewski \cite{slonczewski1996current,slonczewski2002currents} and Berger \cite{berger2001new}.
The model describes a current flow perpendicular to a layer system consisting of a fixed magnetic layer, a nonmagnetic layer and a free magnetic layer.
The current is assumed to pick up its spin polarization in the fixed layer and to exert a torque on the magnetization in the free layer.
While this model is successfully applied to MRAM like structures, it is not suited for the description of current driven domain-wall motion, where no magnetic fixed layer is involved.

A more general model to spin polarized currents introduces a spin-accumulation field that is bidirectionally coupled to the magnetization \cite{zhang2002mechanisms,shpiro2003self}.
The explicit computation of the spin-accumulation can be avoided by neglecting spin-diffusion effects and treating the spin-accumulation in an adiabatic fashion \cite{zhang2004roles}.
This simplified model is able to describe domain-wall motion, but fails in the description of multi-layer structures.

In this work, we consider a spin-diffusion model that represents a more general approach to describe the interaction between spin current and magnetization in a three-dimensional space and time regime.

The model is compared to the results of the simplified models of Slonczewski~\cite{slonczewski1996current,slonczewski2002currents} and Zhang and Li~\cite{shpiro2003self}.
For the discretization, we implement a finite-element scheme that was introduced and mathematically analyzed in our previous work \cite{abert2014spin}.
In numerical simulations, we underline that our model generalizes and combines those of~\cite{slonczewski1996current,slonczewski2002currents,zhang2002mechanisms,shpiro2003self,zhang2004roles} in the sense that the same results are also obtained with the more general model, demonstrating the capability to describe and predict more complex features that depend on an inhomogenous spin current distribution. 

\section*{Model}\label{sec:model}
Let $\omega\subset\mathbb{R}^3$ be some ferromagnetic material.
The magnetization dynamics is described by the Landau-Lifshitz-Gilbert equation
\begin{subequations}\label{eqn:llg}
\begin{align}
  \frac{\partial \boldsymbol{m}}{\partial t} =
  - \gamma \boldsymbol{m} \times (\boldsymbol{h}_\text{eff} + \frac{c}{\mu_0} \boldsymbol{s})
  + \alpha \boldsymbol{m} \times \frac{\partial \boldsymbol{m}}{\partial t}
  \quad\text{in }\omega\times\{t>0\},
\end{align}
subject to initial and boundary conditions
\begin{align}
  \boldsymbol{m}(0) = \boldsymbol{m}^0\text{ in }\omega,
  \quad
  \frac{\partial\boldsymbol{m}}{\partial\boldsymbol{n}}=0 \text{ on }\partial\omega
\end{align}
\end{subequations}
where $\boldsymbol{m}$ is the normalized magnetization, $\gamma$ is the gyromagnetic ratio, and $\alpha$ is the Gilbert damping.  The effective field $\boldsymbol{h}_\text{eff}$ is given by the negative variational derivative of the total free energy $U$
\begin{equation}
  \boldsymbol{h}_\text{eff} = - \frac{1}{\mu_0 M_\text{s}} \frac{\delta U}{\delta \boldsymbol{m}},
  \label{eqn:heff}
\end{equation}
where $\mu_0$ is the magnetic constant and $M_\text{s}$ is the saturation magnetization.
Contributions to the effective field usually include the exchange field given by
\begin{equation}
  \boldsymbol{h}_\text{exch} = \frac{2 A}{\mu_0 M_\text{s}} \Delta \boldsymbol{m}
\end{equation}
with the material dependent exchange constant $A$ and the demagnetization field $\boldsymbol{h}_\text{demag} = - \boldsymbol{\nabla} u$ generated by a magnetic region $\omega$ with the potential $u$ defined by
\begin{alignat}{2}
  \Delta u &= M_\text{s} \; \boldsymbol{\nabla} \cdot \boldsymbol{m} &&\quad\text{in}\quad \omega \label{eqn:demag_first} \\
  \Delta u &= 0                                        &&\quad\text{in}\quad \mathbb{R}^3 \setminus \overline\omega
\intertext{with jump and boundary conditions}
  \left[ u \right]_{\partial \omega} &= 0 &&\quad\text{on}\quad\partial\omega\\
  \left[ \frac{\partial u}{\partial \boldsymbol{n}} \right]_{\partial \omega} &= - M_\text{s} \; \boldsymbol{m} \cdot \boldsymbol{n}&&\quad\text{on}\quad\partial\omega \\
  u(\boldsymbol{x}) &= \mathcal{O}(1/|\boldsymbol{x}|) &&\quad \text{for} \quad |\boldsymbol{x}| \rightarrow \infty
  \label{eqn:demag_last}
\end{alignat}
where $\boldsymbol{n}$ is the outer unit normal of $\omega$ and, e.g., $\boldsymbol{m}\cdot\boldsymbol{n}$ is the Euclidean scalar product.
Further possible contributions are the external field $\boldsymbol{h}_\text{ext}$ and the anisotropy field $\boldsymbol{h}_\text{aniso}$ whose analytical expression depends on the lattice structure of the magnetic material.
The vector field $\boldsymbol{s}$ in \eqref{eqn:llg} is the spin-accumulation generated by a spin current $\boldsymbol{J}_\text{s}$ and $c$ is the corresponding coupling constant.
Let $\Omega\subset\mathbb{R}^3$ be some conducting region with $\omega\subset\Omega$.
According to \cite{garcia2007spin}, the spin-accumulation obeys the equation of motion
\begin{subequations}\label{eqn:spin_diffusion}
\begin{equation}
  \frac{\partial \boldsymbol{s}}{\partial t} =
  - \boldsymbol{\nabla} \cdot \boldsymbol{J}_\text{s}
  - 2 D_0 \left[
    \frac{\boldsymbol{s}}{\lambda_\text{sf}^2}
    + \frac{\boldsymbol{s} \times \boldsymbol{m}}{\lambda_\text{J}^2}
  \right]
  \quad\text{in }\Omega\times\{t>0\},
\end{equation}
subject to initial and boundary conditions
\begin{align}
  \boldsymbol{s}(0) = \boldsymbol{s}^0\text{ in }\Omega,
  \quad
  \frac{\partial\boldsymbol{s}}{\partial\boldsymbol{n}}=0 \text{ on }\partial\Omega
\end{align}
\end{subequations}
where $D_0$ is the diffusion constant, $\lambda_\text{sf}$ is related to the spin-flip relaxation time $\tau_\text{sf}$ by $\lambda_\text{sf} = \sqrt{2 D_0 \tau_\text{sf}}$ and $\lambda_\text{J} = \sqrt{ 2 h D_0 / J_\text{ex}}$ with $h$ being Planck's constant and $J_\text{ex}$ being the exchange integral.
For some given current density $\boldsymbol{J}_\text{e}$, the matrix-valued spin current $\boldsymbol{J}_\text{s}$ is given by
\begin{equation}
  \boldsymbol{J}_\text{s} = 
  \frac{\beta \mu_\text{B}}{e} \boldsymbol{m} \otimes \boldsymbol{J}_\text{e}
  - 2 D_0 \left[
    \boldsymbol{\nabla} \boldsymbol{s}
    - \beta \beta' \boldsymbol{m} \otimes \left( (\boldsymbol{\nabla}\boldsymbol{s})^T \boldsymbol{m} \right)
  \right]
  \label{eqn:spin_current}
\end{equation}
where $\mu_\text{B}$ is the Bohr magneton, $e$ is the elementary charge and $\beta$ and $\beta'$ are dimensionless polarization parameters.
Here, $\boldsymbol{\nabla} \boldsymbol{s}=(\partial\boldsymbol{s}_i/\partial x_j)_{i,j=1}^3\in\mathbb{R}^{3\times3}$ is the Jacobian of $\boldsymbol{s}$ and
$\boldsymbol{a} \otimes \boldsymbol{b} = \boldsymbol{a} \boldsymbol{b}^T\in\mathbb{R}^{3\times3}$ is the outer product.

\begin{figure}
  \centering
  \includegraphics{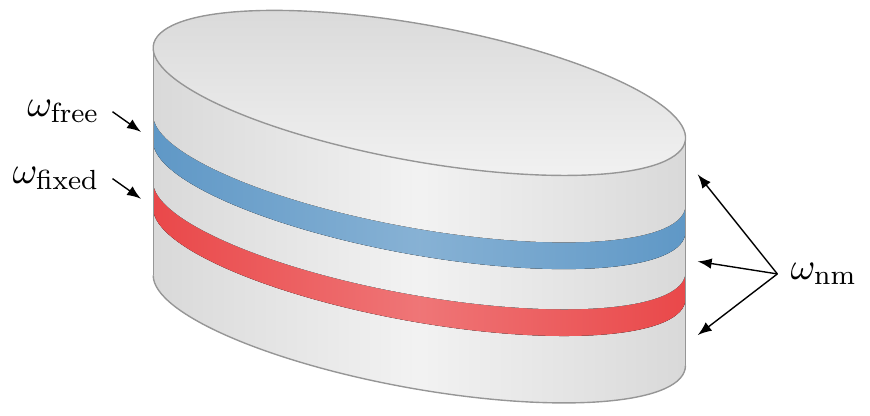}
  \caption{
    Typical multi-layer structure for a spin-torque nano-pillar with $\omega=\omega_\text{fixed}\cup\omega_\text{free}$ and $\Omega = \omega\cup\omega_\text{nm}$.
    A magnetic fixed layer $\omega_\text{fixed}$ is separated from a magnetic free layer $\omega_\text{free}$ by a nonmagnetic spacer layer.
    The system is completed by two leads made of nonmagnetic material.
  }
  \label{fig:sample}
\end{figure}
Figure~\ref{fig:sample} shows a typical multi-layer structure for the investigation of spin-torque effects.
A magnetic fixed layer $\omega_\text{fixed}$ is separated from a magnetic free layer $\omega_\text{free}$ by a nonmagnetic layer $\omega_\text{nm}$.
Additional nonmagnetic layers above and beneath the magnetic layers serve as electrical contacts.
The Landau-Lifshitz-Gilbert equation \eqref{eqn:llg} is solved on the magnetic regions $\omega = \omega_\text{fixed} \cup \omega_\text{free}$ only.
The spin-accumulation is solved on the entire domain $\Omega = \omega \cup \omega_\text{nm}$.

\section*{Algorithm}\label{sec:algorithm}
\def\m{\boldsymbol{m}}
\def\s{\boldsymbol{s}}
\def\v{\boldsymbol{v}}
\def\A{\boldsymbol{A}}
\def\B{\boldsymbol{B}}
\def\R{\mathbb{R}}
The model introduced in the preceding section is discretized by an algorithm, we proposed and analyzed mathematically in~\cite{abert2014spin}. The vector fields $\boldsymbol{m}$, $\boldsymbol{s}$, $\boldsymbol{v}=\partial\m/\partial t$, and test functions $\boldsymbol{w}$, $\boldsymbol{\zeta}$ are discretized with piecewise affine, globally continuous functions constructed on a tetrahedral tessellation of the domain $\Omega$.
Material constants that differ from subdomain to subdomain, e.g., the diffusion constant $D_0$, are discretized by piecewise constant functions. A splitting scheme is used to decouple the Landau-Lifshitz-Gilbert equation \eqref{eqn:llg} from the spin-accumulation equation \eqref{eqn:spin_diffusion}.

Suppose that the magnetization $\m^{k}$ at the $k$-th timestep is known. Then, we use the tangent-plane scheme of~\cite{alouges} to compute $\m^{k+1}$ at the next time-step, where the essential idea is to discretize $\partial\m/\partial t$ at the $k$-th timestep by some $\v^k$ which mimics the continuous orthogonality $\m\cdot(\partial\m/\partial t) = 0$.
 Following~\cite{akt,ghps,multiscale}, all terms but the exchange term are treated explicitly. This also includes the spin-accumulation $\boldsymbol{s}$. Based on $\m^{k+1}$ and $\s^{k}$, we then use the implicit Euler scheme to compute the spin-accumulation $\boldsymbol{s}^{k+1}$ at the next time-step.
The complete algorithm for a timestep size $\tau>0$ reads as follows, where we suppose that $\m^k$ and $\s^k$ are already known:
\begin{enumerate}
  \item
    Consider the discrete tangent space $\mathcal{T}_{\m^k}$ to the known magnetization $\boldsymbol{m}^k$ defined by
    \begin{equation}
      \mathcal{T}_{\m^k} = \{ \v \text{ piecewise affine and globally continuous}: \boldsymbol{v} \cdot \boldsymbol{m}^k = 0 \text{ at all nodes}\}.
    \end{equation}
    Find $\boldsymbol{v}^{k} \in \mathcal{T}_{\m^k}$ such that, for all $\boldsymbol{w} \in \mathcal{T}_{\m^k}$, it holds
\begin{multline}
  \int_{\omega} \left( \alpha \boldsymbol{v}^{k} + \boldsymbol{m}^{k} \times \boldsymbol{v}^{k} \right) \cdot \boldsymbol{w} \;\text{d}\boldsymbol{x}
  + \frac{2 \gamma A \tau}{\mu_0 M_\text{s}} \int_{\omega} \boldsymbol{\nabla} \boldsymbol{v}^{k} : \boldsymbol{\nabla} \boldsymbol{w} \;\text{d}\boldsymbol{x}
  =\\
  =- \frac{2 \gamma A}{\mu_0 M_\text{s}} \int_{\omega} \boldsymbol{\nabla} \boldsymbol{m}^{k} : \boldsymbol{\nabla} \boldsymbol{w} \;\text{d}\boldsymbol{x}
  + \gamma \int_{\omega} (\boldsymbol{h}_\text{explicit} + \frac{c}{\mu_0}\,\s^k)\cdot \boldsymbol{w} \;\text{d}\boldsymbol{x},
  \label{eqn:weak_llg}
\end{multline}
where $\boldsymbol{A} : \boldsymbol{B}=\sum_{i,j=1}^3\A_{ij}\B_{ij}$ is the Frobenius inner product of two matrices $\A,\B\in\R^{3\times3}$.
While the exchange field ist integrated implicitly, all other effective-field contributions are summarized in $\boldsymbol{h}_\text{explicit}$ and integrated explicitly.

\item Define the piecewise affine and globally continuous approximation $\boldsymbol{m}^{k+1}$ by
\begin{equation}
  \boldsymbol{m}^{k+1} = \frac{\boldsymbol{m}^k + \tau \boldsymbol{v}^k}{|\boldsymbol{m}^k + \tau \boldsymbol{v}^k|}\text{ nodewise}.
\end{equation}

\item Find $\boldsymbol{s}^{k+1}$ such that, for all $\boldsymbol{\zeta}$, it holds
\begin{multline}
  \int_{\Omega} 
  \frac{\boldsymbol{s}^{k+1} - \boldsymbol{s}^{k}}{\tau}\cdot \boldsymbol{\zeta} \;\text{d}\boldsymbol{x}
  \;+\; 2 D_0 \, a(\boldsymbol{s}^{k+1}, \boldsymbol{\zeta})
  =\\
  =\frac{\beta \mu_\text{B}}{e} \int_{\omega} \left[ \boldsymbol{m}^{k+1} \otimes \boldsymbol{J}_\text{e}^{k+1} \right] : \boldsymbol{\nabla} \boldsymbol{\zeta} \;\text{d}\boldsymbol{x}
  - \frac{\beta \mu_\text{B}}{e} \int_{\partial \Omega \cap \partial \omega} (\boldsymbol{J}_\text{e}^{k+1} \cdot \boldsymbol{n}) (\boldsymbol{m}^{k+1} \cdot \boldsymbol{\zeta}) \;\text{d}\boldsymbol{x},
  \label{eqn:weak_spin_diffusion}
\end{multline}
 where $\boldsymbol{n}$ is the outer normal of $\omega$
and
\begin{multline}
  a(\boldsymbol{\zeta_1}, \boldsymbol{\zeta_2}) =
  \frac{1}{\lambda_\text{sf}^2} \int_{\Omega} \boldsymbol{\zeta}_1 \cdot \boldsymbol{\zeta}_2 \;\text{d}\boldsymbol{x}
  + \int_{\Omega} \boldsymbol{\nabla} \boldsymbol{\zeta}_1 : \boldsymbol{\nabla} \boldsymbol{\zeta}_2 \;\text{d}\boldsymbol{x}
  \\
  - \beta \beta' \int_{\omega} \left[ \boldsymbol{m}^{k+1} \otimes \left((\boldsymbol{\nabla} \boldsymbol{\zeta}_1)^T \boldsymbol{m}^{k+1} \right) \right] : \boldsymbol{\nabla} \boldsymbol{\zeta}_2 \;\text{d}\boldsymbol{x}
  + \frac{1}{\lambda_\text{J}^2} \int_{\omega} \left( \boldsymbol{\zeta}_1 \times \boldsymbol{m}^{k+1} \right) \cdot \boldsymbol{\zeta}_2 \;\text{d}\boldsymbol{x}.
\end{multline}
\item Update the timestep and iterate.
\end{enumerate}
While our preceding work~\cite{abert2014spin} did not include any numerical simulations with the proposed time-marching scheme, we proved that the coupled system of LLG with spin-diffusion admits weak solutions and that the proposed integrator converges to these in an appropriate sense. With the present work, we aim to provide empirical evidence for the physical relevance of the proposed model and the numerical results that are obtained by the described discretization.

\section*{Implementation}\label{sec:implementation}
The presented algorithm is implemented within the finite-element micromagnetic code magnum.fe \cite{abert2013magnum,magnum_website}.
magnum.fe uses FEniCS \cite{logg2012automated} for the assembly and solution of the finite-element systems.
The linear systems arising from discretization of \eqref{eqn:weak_llg} and \eqref{eqn:weak_spin_diffusion} are solved iteratively by a GMRES solver.
The demagnetization-field problem is solved by a hybrid finite-element boundary-element method \cite{fredkin1990hybrid}, where our implementation couples FEniCS to the boundary-element library BEM++ \cite{bem++}.

\section*{Numerical Experiments}\label{sec:experiments}
As a first test the time development of the spin diffusion for a fixed magnetization configuration is computed.
An ellipsoidal multi-layer structure as pictured in Fig.~\ref{fig:sample} with axis lengths $130 \times 70\,\text{nm}$ and layer thicknesses $(5,2,3,2,5)\,\text{nm}$ is considered.
In the following the coordinate system is chosen such that the long and short axes of the ellipse align with the $x$ and $y$-axis respectively.
The diffusion constant in the magnetic layers $\omega_\text{free}$ and $\omega_\text{fixed}$ is set to $D_0 = 1\times10^{-3}\,\text{m$^2$/s}$.
In the nonmagnetic region the diffusion constant is chosen as $D_0 = 5\times10^{-3}\,\text{m$^2$/s}$.
The remaining constants are $\lambda_\text{sf} = 10\,\text{nm}$, $\lambda_\text{J} = 2\,\text{nm}$, $\beta = 0.9$, $\beta' = 0.8$, and $c = 3.125 \times 10^{-3}\,\text{N/A$^2$}$.
The magnetization $\boldsymbol{m}$ is initialized homogeneously parallel to the $x$-axis in the fixed layer and parallel to $y$-axis in the free layer.
Equation~\eqref{eqn:weak_spin_diffusion} is then solved repeatedly for a homogeneous current density perpendicular to the layers $\boldsymbol{J}_\text{e} = (0,0,10^{11})\,\text{A/m$^2$}$, a time step $\tau = 1\,\text{fs}$ and an initial spin-accumulation of $\boldsymbol{s} = 0$.

\begin{figure}
  \subfloat[]{
    \includegraphics{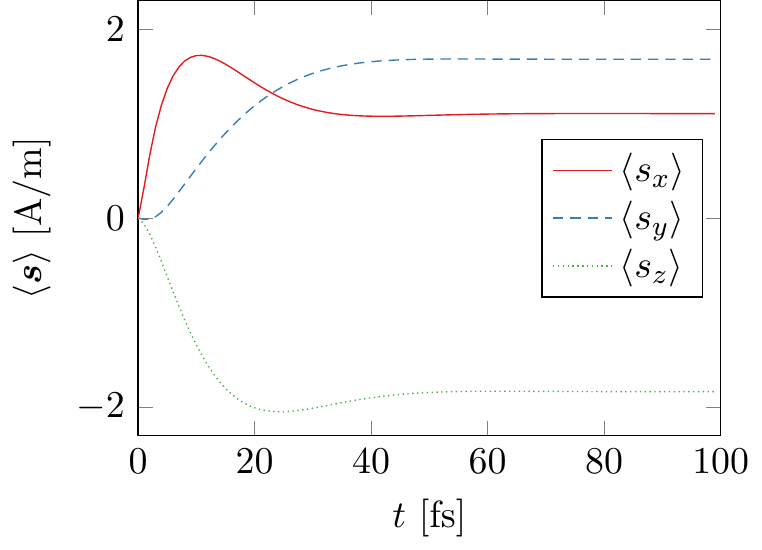}
    \label{fig:relax_s}
  }
  \subfloat[]{
    \includegraphics{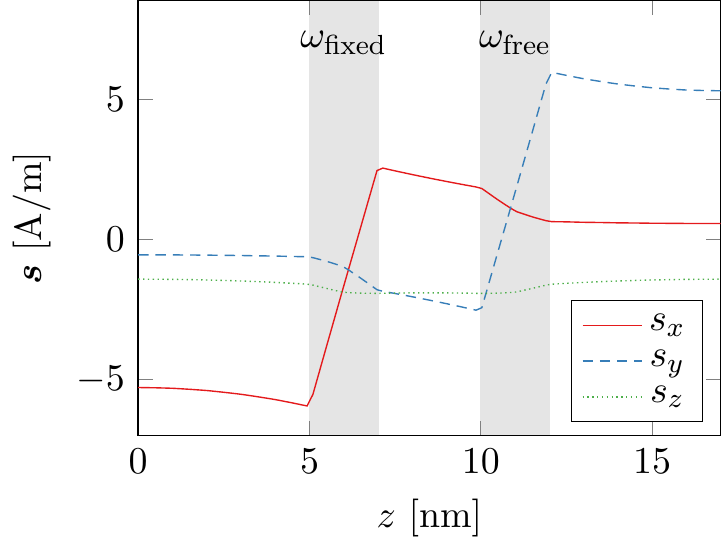}
    \label{fig:s_on_line}
  }
  \caption{
    spin-accumulation in a magnetic multi-layer structure.
    The magnetization is homogeneous and in-plane within the magnetic regions.
    (a) Spatially averaged components of the spin diffusion in the free layer during relaxation.
    (b) Spin diffusion in equilibrium along the the $z$-axis of the multi-layer.
  }
  \label{fig:s}
\end{figure}
Figure~\ref{fig:relax_s} shows the simulation results.
The system relaxes within approximately $70\,\text{fs}$.
This time scale is 2 orders of magnitude below the typical reaction time of the magnetization $\boldsymbol{m}$.
Hence the magnetization dynamics may be accurately described by assuming the spin-accumulation to be in equilibrium at any times \cite{zhang2002mechanisms}.
However, the presented integration scheme for the spin-accumulation is able to handle very large time steps due to its implicit nature and the linearity of the problem.
In fact it shows that even for an initial spin-accumulation of $\boldsymbol{s} = 0$ two integration steps with $\tau = 1\,\text{ps}$ are sufficient to obtain the equilibrium state accurately.
When moving from one magnetization configuration to the next during time integration of the Landau-Lifshitz-Gilbert equation, a single integration step of the spin-accumulation yields the equilibrium state.
The presented algorithm is thus well suited for investigating the time evolution of both the spin diffusion using time steps $\tau \approx 1\,\text{fs}$ as well as the magnetization using time steps $\tau \approx 1\,\text{ps}$.

Figure~\ref{fig:s_on_line} shows the equilibrium spin-accumulation $\boldsymbol{s}$ on the $z$-axis in the middle of the multi-layer structure.
From \eqref{eqn:spin_diffusion} it is clear, that the source of the spin-accumulation is the spatial change of the magnetization $\boldsymbol{m}$.
Since the magnetization is homogeneous within the magnetic regions, the spin-accumulation is expected to be generated at the layer interfaces and diffuse into the regions which is well reflected by the simulation results.

\subsection*{Comparison to Model by Slonczewski}\label{sec:slonczewski}
An established model for the description of spin-torque effects in magnetic multi-layers was introduced by Slonczewski in \cite{slonczewski2002currents}.
The fixed layer in this model is homogeneously magnetized and fixed in the sense that the magnetization does not change in time.
The electric current $\boldsymbol{J}_\text{e}$ is assumed to pick up its spin polarization in the fixed layer and then interact with the free layer by an additional term to the Landau-Lifshitz-Gilbert equation
\begin{equation}
  \boldsymbol{N} = \eta(\theta) \frac{\hbar}{2e} \frac{J_\text{e}}{d} \boldsymbol{m} \times \left( \boldsymbol{m} \times \boldsymbol{M} \right).
  \label{eqn:slonczewski_torque}
\end{equation}
Here $\boldsymbol{M}$ denotes the normalized magnetization of the fixed layer, $d$ is the thickness of the free layer and $\theta$ is the angle between the magnetization in the fixed layer and the magnetization with $\cos(\theta) = \boldsymbol{M} \cdot \boldsymbol{m}$.
The additional contribution $\boldsymbol{N}$ is added to the right-hand side of the Landau-Lifshitz-Gilbert equation \eqref{eqn:llg} without spin-accumulation $\boldsymbol{s}$.
The angular coefficient $\eta(\theta)$ for structures with equal height fixed and free layers is given by
\begin{equation}
  \eta(\theta) = \frac{q}{A + B \cos(\theta)}
  \label{eqn:slonczewski_eta}
\end{equation}
where $q$, $A$, and $B$ are constants that depend on the geometry and the material parameters of the system.
\begin{figure}
  \centering
  \subfloat[]{
    \includegraphics{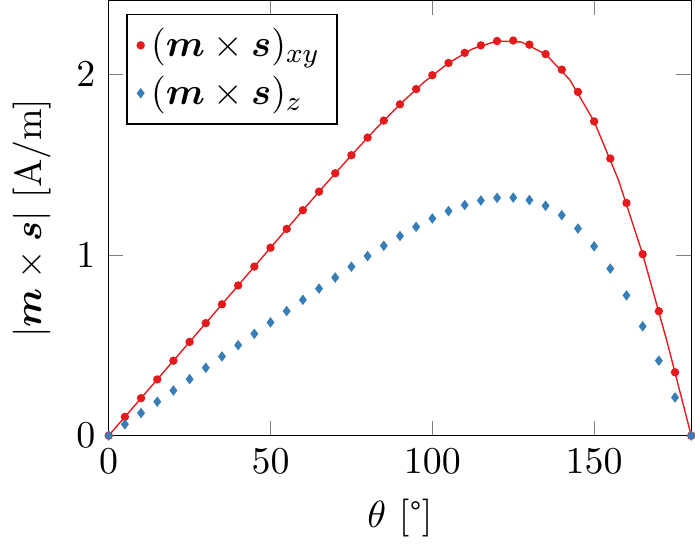}
    \label{fig:slonczewski_symmetric}
  }
  \subfloat[]{
    \includegraphics{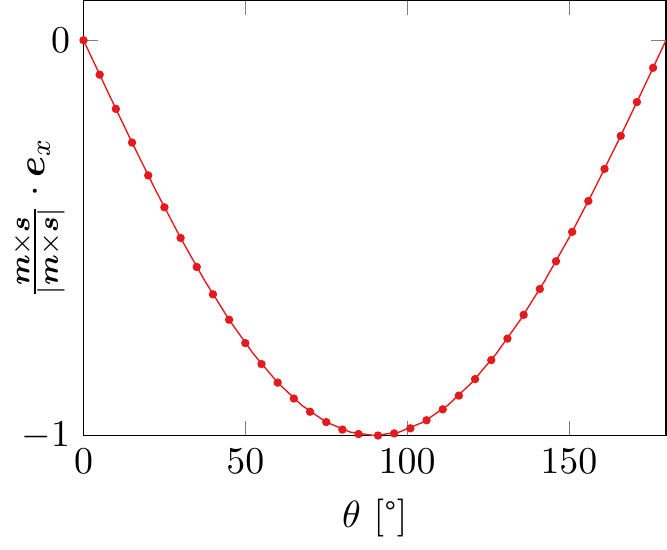}
    \label{fig:slonczewski_symmetric_angle}
  }
  \caption{
    spin-torque $\boldsymbol{m} \times \boldsymbol{s}$ in a magnetic multi-layer structure with layer thicknesses $(5, 2, 3, 2, 5)\,\text{nm}$ and $\lambda_\text{J} = 2\,\text{nm}$.
    The magnetization in the fixed layer and the free layer is homogeneous, in-plane, and tilted by $\theta$.
    The marks represent values computed with the spin diffusion model and the lines show a fit with the model of Slonczewski.
    (a) In-plane and out-of-plane component of the equilibrium spin-torque.
    (b) In-plane angle of the spin-torque with respect to the magnetization direction in the fixed layer.
  }
\end{figure}
Consider the previously described ellipsoidal multi-layer structure with the magnetization being homogeneous and in-plane in both the fixed layer and the free layer.
In this case the additional torque term in the free layer \eqref{eqn:slonczewski_torque} is expected to be in-plane  with a magnitude given by
\begin{equation}
  N \propto \sin(\theta)\frac{q}{A + B \cos(\theta)}.
  \label{eqn:slonczewski_magnitude}
\end{equation}
Figure~\ref{fig:slonczewski_symmetric} shows the magnitude of the torque term $\boldsymbol{m} \times \boldsymbol{s}$ as computed by the spin-accumulation model.
The result is split into the in-plane magnitude $(\boldsymbol{m} \times \boldsymbol{s})_{xy}$ and the out-of-plane magnitude $(\boldsymbol{m} \times \boldsymbol{s})_{z}$.
Furthermore the in-plane magnitude is fitted with the magnitude as predicted by Slonczewski \eqref{eqn:slonczewski_magnitude}.
Since the system is overdefined for fitting, $q$ is set to 1 and $A$ and $B$ are chosen as fitting parameters.
As seen in Fig.~\ref{fig:slonczewski_symmetric} the fit is very accurate.
The relative fitting error for both $A$ and $B$ is well below $10^{-6}$.
Not only the in-plane magnitude, but also the in-plane angle of the torque term shows excellent accordance to the spin-torque model of Slonczewski as depicted in Fig.~\ref{fig:slonczewski_symmetric_angle}.
However, the spin-accumulation model predicts a non-neglible out-of-plane torque that cannot be explained by the simplified model.

\begin{figure}
  \centering
  \subfloat[]{
    \includegraphics{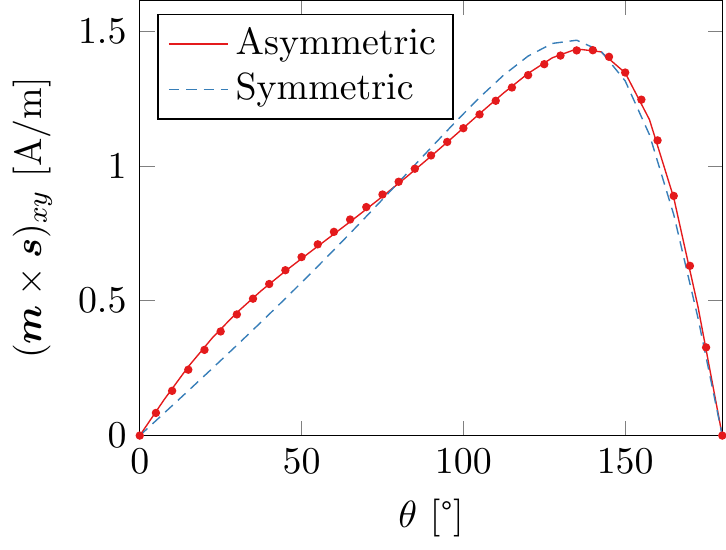}
    \label{fig:slonczewski_asymmetric}
  }
  \subfloat[]{
    \includegraphics{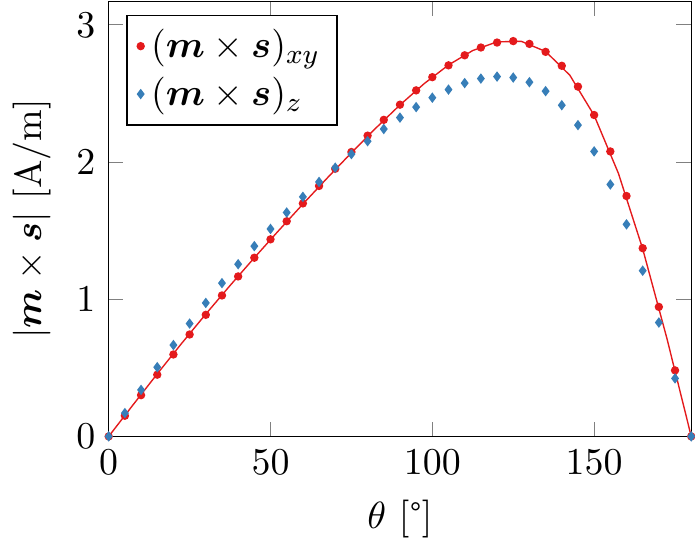}
    \label{fig:slonczewski_asymmetric_lambda_2}
  }
  \caption{
    spin-torque $\boldsymbol{m} \times \boldsymbol{s}$ in a magnetic multi-layer structure with layer thicknesses $(5, 10, 3, 2, 5)\,\text{nm}$.
    The marks represent values computed with the spin diffusion model and the lines show a fit with the model of Slonczewski.
    (a) In-plane spin-torque for $\lambda_\text{J} = 1\,\text{nm}$. The data is fitted with both a symmetric and an asymmetric model.
    (b) In-plane and out-of-plane component of the spin-torque for $\lambda_\text{J} = 2\,\text{nm}$.
  }
\end{figure}
As pointed out in \cite{xiao2004boltzmann} the expression for the angular coefficient \eqref{eqn:slonczewski_eta} only holds for the symmetric case of equal height fixed and free layer.
For the general case the angular coefficient reads
\begin{equation}
  \eta(\theta) = \frac{q_+}{A + B \cos(\theta)}
               + \frac{q_-}{A - B \cos(\theta)}.
  \label{eqn:slonczewski_eta_asymmetric}
\end{equation}
Figure~\ref{fig:slonczewski_asymmetric} shows the computed in-plane torque for $\lambda_\text{J} = 1\,\text{nm}$ together with fitted curves for the symmetric model \eqref{eqn:slonczewski_eta} and the asymmetric model \eqref{eqn:slonczewski_eta_asymmetric} respectively.
The computed torque is well fitted by the asymmetric model with relative errors below $2\times10^{-4}$, while the symmetric model obviously fails in the correct description of the asymmetric problem.
However, again the spin-diffusion model predicts a $z$-component of the torque that is not described by the spin-torque model of Slonczewski.
The ratio of in-plane torque to out-of-plane torque for a given structure is largely influenced by the choice of $\lambda_\text{J}$.
Choosing large values for $\lambda_\text{J}$ leads to high out-of-plane components of the torque, see Fig.~\ref{fig:slonczewski_asymmetric_lambda_2}.

The coefficients $A$, $B$, $q$, $q_+$, and $q_-$ depend, among other things, on the geometry of the system.
In the following the dependence of these constants on the thickness of the nonmagnetic layer and the fixed layer is investigated in detail.
A physically motivated choice for the right-hand-side coefficients in \eqref{eqn:slonczewski_eta_asymmetric} is introduced in \cite{xiao2004boltzmann}
\begin{align}
  q_\pm &= \frac{1}{2} \left[
    P_\text{L} \Lambda_\text{L}^2 \sqrt{\frac{\Lambda_\text{R}^2 + 1}{\Lambda_\text{L}^2 + 1}}
    \pm
    P_\text{R} \Lambda_\text{R}^2 \sqrt{\frac{\Lambda_\text{L}^2 + 1}{\Lambda_\text{R}^2 + 1}}
  \right]\\
  A &= \sqrt{ (\Lambda_\text{L}^2 + 1) (\Lambda_\text{R}^2 + 1)}\\
  B &= \sqrt{ (\Lambda_\text{L}^2 - 1) (\Lambda_\text{R}^2 - 1)}
\end{align}
where $\Lambda_\text{L}$, $\Lambda_\text{R}$, $P_\text{L}$, and $P_\text{R}$ describe the properties of the fixed and the free layer respectively.
For symmetric structures, i.e. $\Lambda_\text{L} = \Lambda_\text{R}$ and $P_\text{L} = P_\text{R}$, \eqref{eqn:slonczewski_eta_asymmetric} reduces to
\begin{equation}
  \eta(\theta) = \frac{P \Lambda^2}{(\Lambda^2+1) + (\Lambda^2-1) \cos(\theta)}.
\end{equation}
\begin{figure}
  \centering
  \subfloat[]{
    \includegraphics{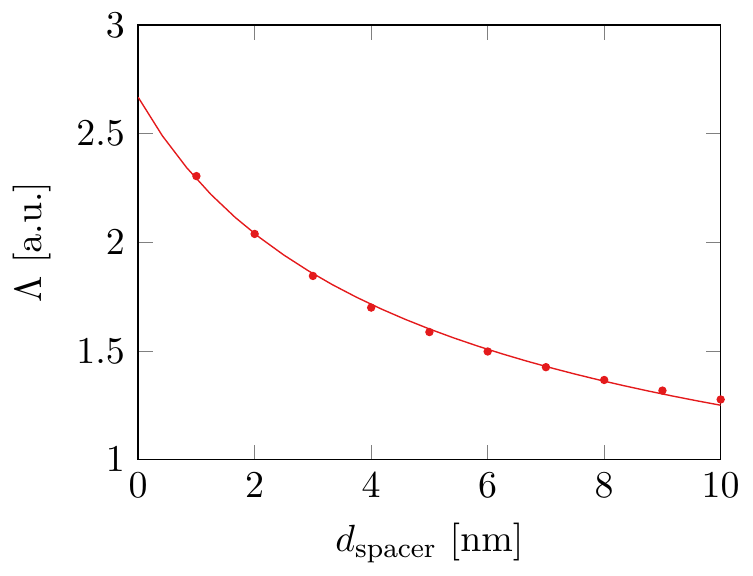}
    \label{fig:lambda_spacer}
  }
  \subfloat[]{
    \includegraphics{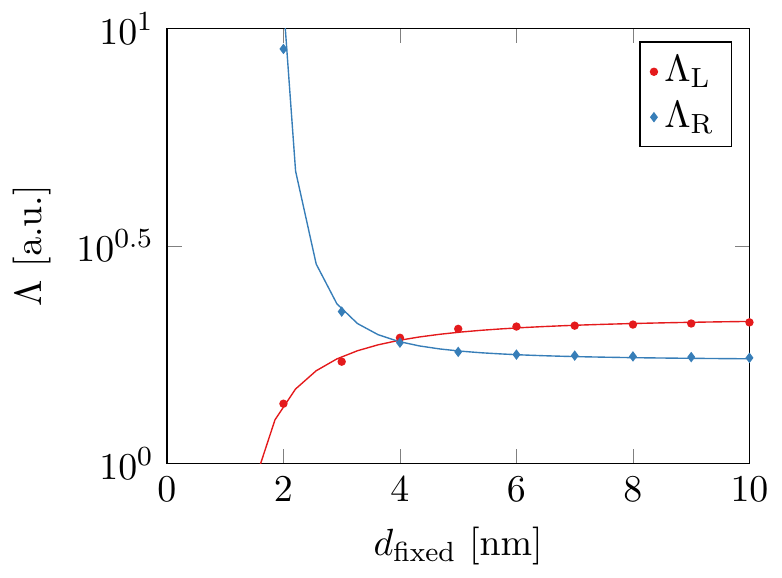}
    \label{fig:lambda_fixed_layer}
  }
  \caption{
    Dependence of the fitting parameter $\Lambda$ on the geometry of a multi-layer structure with layer thicknesses $(5,2,3,2,5)\,\text{nm}$.
    (a) Variation of the spacer thickness, i.e. $(5,2,d_\text{spacer},2,5)\,\text{nm}$.
    (b) Variation of the fixed-layer thickness, i.e. $(5,d_\text{fixed},3,2,5)\,\text{nm}$.
  }
\end{figure}
Figure~\ref{fig:lambda_spacer} shows the dependence of $\Lambda$ from the thickness of the spacer layer in a symmetric multi-layer structure.
The simulation results are well fitted with a shifted reciprocal square root $\Lambda(d) \propto 1 / \sqrt{\Delta + d}$.
The dependence of $\Lambda_\text{L}$ and $\Lambda_\text{R}$ on the thickness of the fixed layer in an asymmetric structure is depicted in Fig.~\ref{fig:lambda_fixed_layer}.
Both coefficients are well fitted by a shifted reciprocal square function $\Lambda_\text{L/R}(d) = X + Y / (\Delta + d)$ with $X$, $Y$, and $\Delta$ being fit parameters.

The model of Slonczewski provides a straightforward way to deal with multi-layer structures.
However, it is obviously not suited for the description of spin-torque effects in smoothly variing magnetization configurations since it requires a separated fixed layer to generate the spin polarization of the current.

\subsection*{Comparison to Model by Zhang and Li}\label{sec:zhang}
An alternative spin-torque model was introduced by Zhang and Li in \cite{zhang2004roles}.
The model of Zhang and Li can be derived from the spin-diffusion model \eqref{eqn:spin_diffusion} by introducing the following simplifications:
The spin diffusion $\boldsymbol{s}$ is assumed to vary little in space and thus terms involving $\boldsymbol{\nabla} \boldsymbol{s}$ are omitted.
The electric current is assumed to be divergence free within the simulated regions, i.e. $\boldsymbol{\nabla} \cdot \boldsymbol{J}_\text{e} = 0$.
Furthermore the spin-accumulation is assumed to be in equilibrium at all times, i.e. $\partial \boldsymbol{s} / \partial t = 0$.
With these assumptions \eqref{eqn:spin_diffusion} reads
\begin{align}
  \frac{\partial \boldsymbol{s}}{\partial t}
  &=
  - \frac{\beta \mu_\text{B}}{e} \boldsymbol{\nabla} \cdot \left( \boldsymbol{m} \otimes \boldsymbol{J}_\text{e} \right)
  - 2 D_0 \left[
    \frac{\boldsymbol{s}}{\lambda_\text{sf}^2}
    + \frac{\boldsymbol{s} \times \boldsymbol{m}}{\lambda_\text{J}^2}
  \right]
  \\
  &=
  - \frac{\beta \mu_\text{B}}{e}
    (\boldsymbol{J}_\text{e} \cdot \boldsymbol{\nabla}) \boldsymbol{m}
  - 2 D_0 \left[
    \frac{\boldsymbol{s}}{\lambda_\text{sf}^2}
    + \frac{\boldsymbol{s} \times \boldsymbol{m}}{\lambda_\text{J}^2}
  \right]
  \\
  &= 0.
\end{align}
Applying basic vector algebra and $\boldsymbol{m} \cdot \boldsymbol{m} = 1$ yields
\begin{align}
  \boldsymbol{m} \times \boldsymbol{s}
  =
  \frac{1}{1 + \xi^2} \frac{\lambda_\text{J}^2 \beta \mu_\text{B}}{2 D_0 e}
  \Big(
  - \boldsymbol{m} \times [\boldsymbol{m} \times (\boldsymbol{J}_\text{e} \cdot \boldsymbol{\nabla}) \boldsymbol{m}]
  - \xi \boldsymbol{m} \times (\boldsymbol{J}_\text{e} \cdot \boldsymbol{\nabla}) \boldsymbol{m}
  \Big)
  \label{eqn:zhang_torque}
\end{align}
with the degree of non-adiabacity $\xi = \lambda_\text{J}^2 / \lambda_\text{sf}^2$.
The obvious advantage of this method is that the spin-accumulation does not have to be calculated explicitly.
The spin-torque $\boldsymbol{m} \times \boldsymbol{s}$ does only depend on the magnetization $\boldsymbol{m}$ and other known entities and can be added directly to the Landau-Lifshitz-Gilbert equation \eqref{eqn:llg}.

\begin{figure}
  \centering
  \subfloat[]{
    \includegraphics{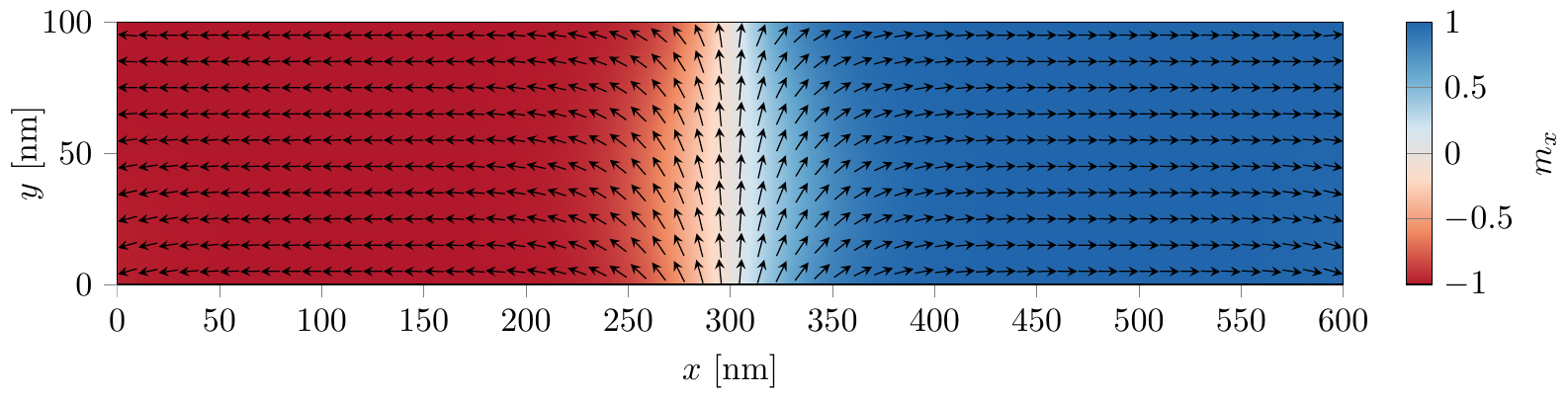}
    \label{fig:domain_wall_m}
  }
  \\
  \subfloat[]{
    \includegraphics{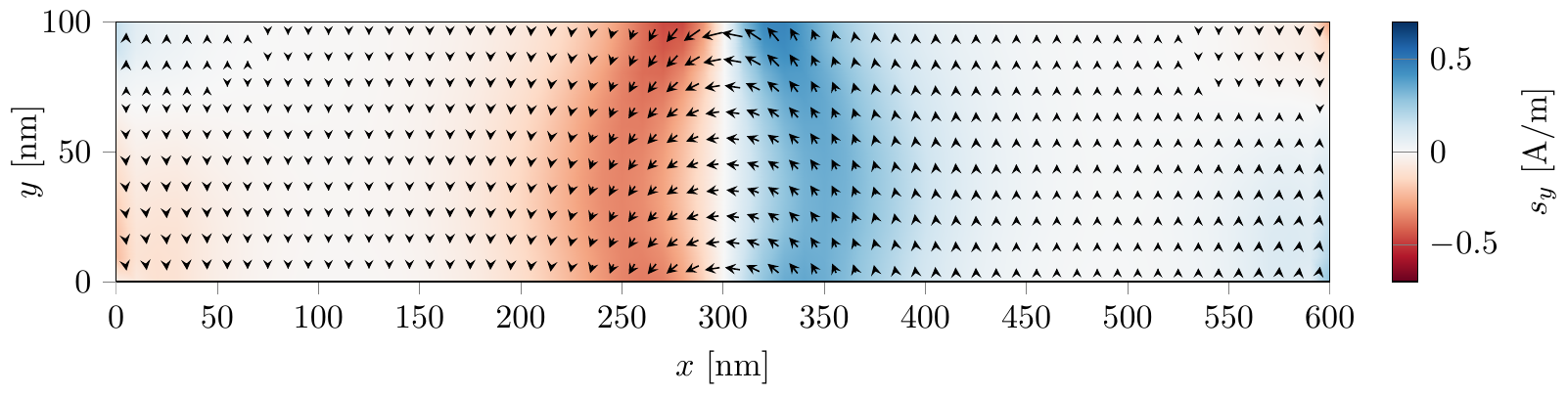}
    \label{fig:domain_wall_s}
  }
  \caption{
    N\'eel domain wall in a thin strip of permalloy with dimensions $600 \times 100 \times 2\,\text{nm}$.
    (a) Magnetization configuration. The $x$-component is color coded.
    (b) Spin-torque. The $y$-component is color coded.
  }
  \label{fig:domain_wall}
\end{figure}
Figure~\ref{fig:domain_wall_m} shows the magnetization configuration $\boldsymbol{m}$ for a N\'eel-wall in a $600 \times 100 \times 2\,\text{nm}$ thin film with the material parameters of permalloy ($M_\text{s} = 8\times10^{5}\,\text{A/m}$, $A=1.3\times10^{-11}\,\text{J/m}$, $K=0$) as computed with magnum.fe.
The corresponding spin-torque $\boldsymbol{m} \times \boldsymbol{s}$ is computed once by relaxation of the spin-diffusion model \eqref{eqn:spin_diffusion}, see Fig.~\ref{fig:domain_wall_s}, and once by the model of Zhang and Li \eqref{eqn:zhang_torque}.
The solution computed by the model of Zhang and Li shows a good agreement to the spin-diffusion solution with a relative $L^2$ error norm below $5 \times 10^{-3}$. 

The assumption of a vanishing gradient of the spin-accumulation $\boldsymbol{\nabla} \boldsymbol{s}$, made in the model by Zhang and Li, is valid within the magnetic regions as discussed in \cite{zhang2004roles}.
However, at the interfaces of magnetic multi-layers this assumption is violated.
Hence the model by Zhang and Li is not suited for the description of such structures.

\subsection*{Standard Problem \#5}
A suitable benchmark for the interplay of magnetization and spin-diffusion dynamics is the micromagnetic standard problem \#5 that was recently proposed by the \textmu Mag group, see \cite{mumag_sp5}.
A thin square of size $100\times100\times10\,\text{nm}$ with material parameters similar to permalloy is prepared in a vortex state.
Then a constant current $J_\text{e} = 10^{12}\,\text{A/m}$ is applied in $x$-direction.
The vortex core is expected to perform a damped rotation around a new equilibrium position.
The spin-torque related costants in the problem definition are tailored to the model by Zhang and Li.
Namely the degree of non-adiabacity $\xi$ is given as well as a factor $b$ that can be expressed in terms of spin-diffusion related constants as
\begin{equation}
  b = \frac{1}{1 + \xi^2} \frac{\gamma c \lambda_\text{J}^2 \beta \mu_\text{B}}{2 D_0 e \mu_0}
\end{equation}
according to \eqref{eqn:zhang_torque}.
The spin-diffusion related parameters are chosen as $D_0 = 10^{-3}\,\text{m/s$^2$}$, $\beta = 0.9$, $\beta' = 0.8$, $\lambda_\text{sf} = 10\,\text{nm}$, $\lambda_\text{J} \approx 2.236\,\text{nm}$, and
$c \approx 3.155 \times 10^{-3}\,\text{N/A$^2$}$.
In fact $\lambda_\text{J}$ and $c$ are chosen such that $\xi= 0.05$ and $bJ = 72.17\,\text{m/s}$ is exactly fulfilled as required.

\begin{figure}
  \subfloat[]{
    \includegraphics{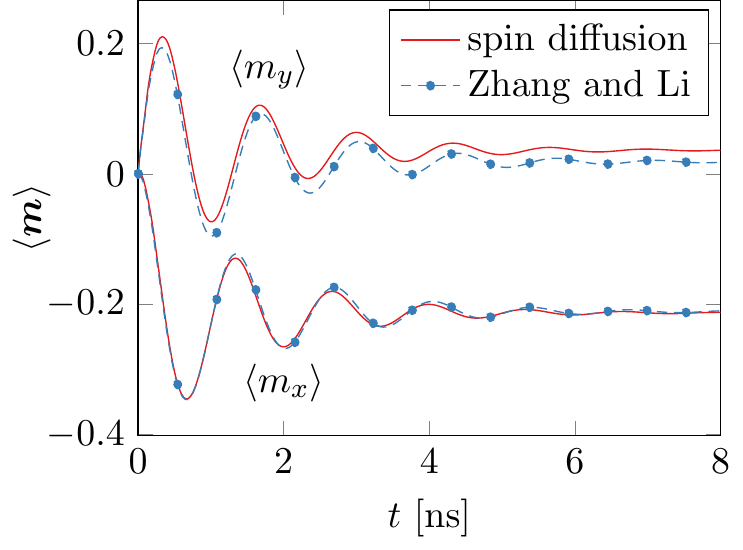}
    \label{fig:sp5_m}
  }
  \subfloat[]{
    \includegraphics{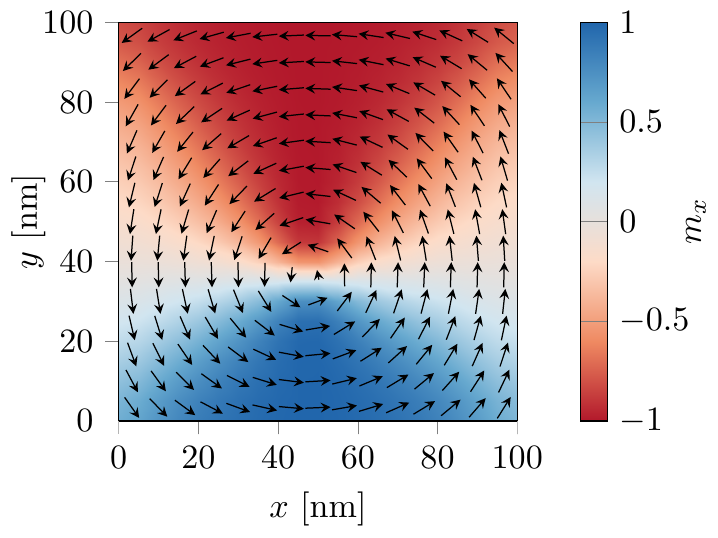}
    \label{fig:sp5_vortex}
  }
  \caption{
    Results for the \textmu Mag standard problem \#5.
    (a) Time evolution of the spatially averaged magnetization compared to a finite-difference simulation according to Zhang and Li.
    (b) New equilibrium position of the magnetization.
  }
  \label{fig:sp5}
\end{figure}
Figure~\ref{fig:sp5} shows the simulation results in comparsion to a reference solution computed with the finite-difference code MicroMagnum \cite{micromagnum} that implements the model by Zhang and Li.
The most notable difference is the shifted $y$-position of the vortex core in equilibrium.
The authors believe that this is a result of the missing $\boldsymbol{\nabla} \boldsymbol{s}$ terms in the model by Zhang and Li.

\subsection*{Switching of a Permalloy multi-layer Structure}
As another test the current induced switching of a magnetic multi-layer structure is simulated.
The geometry considered is ellipsoidal with edge lengths $130 \times 70\,\text{nm}$ and layer thicknesses $(3,10,3,2,3)\,\text{nm}$.
The material parameters in the magnetic layers are similar to those of permalloy, see preceeding section.
In the nonmagnetic layers the diffusion constant is set to $D_0 = 5\times10^{-3}\,\text{m/s$^2$}$. 
Since the size of the ellipsoid is well under the single-domain limit, the energy of the system is minimal if the magnetization is aligned along the long axis of the ellipsoid in the magnetic layers.
The thick and thin magnetic layers are referred to as fixed and free layer respectively, since the energy barrier for magnetic switching in the thick layer is higher than in the thin layer.

\begin{figure}
  \subfloat[]{
    \includegraphics{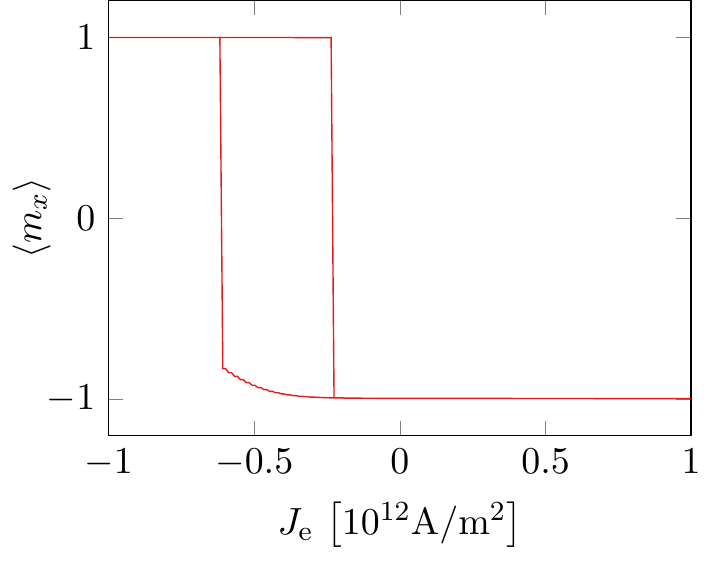}
    \label{fig:switch_hyst}
  }
  \subfloat[]{
    \includegraphics{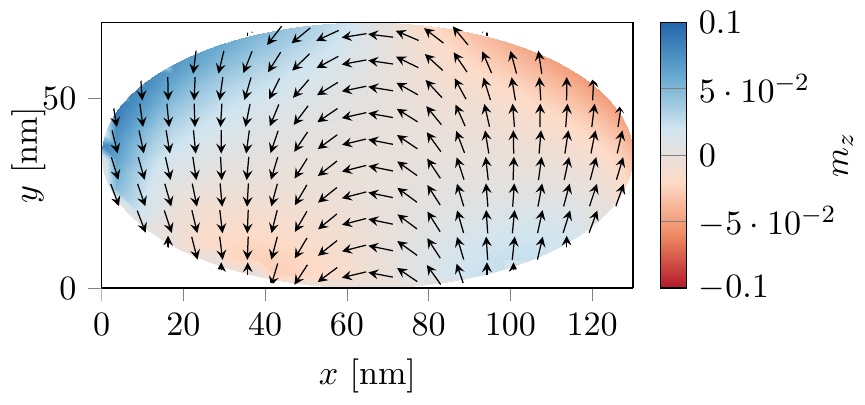}
    \label{fig:switch_m}
  }
  \caption{
    Current hysteresis of a permalloy multi-layer structure.
    (a) Averaged $x$-component of the magnetization in the magnetic free layer.
    (b) Magnetization configuration in the free layer during switching.
  }
  \label{fig:switch}
\end{figure}
By applying a current perpendicular to the layer structure, the magnetization in the free layer can be switched.
Figure~\ref{fig:switch_hyst} shows the hysteresis of the free layer for different currents.
The shift of the curve is an expected consequence resulting from the strayfield coupling  of the magnetic layers.
An antiparallel configuration of the fixed and free layer is energetically favored over a parallel configuration.

Figure~\ref{fig:switch_m} shows the magnetization configuration of the free layer during switching.
Note that the out-of plane component is small compared to the in-plane component, which is a consequence of the shape anisotropy caused by the small thickness of the layer.
The out-of-plane component of the torque as predicted by the spin-diffusion model is compensated by the shape anisotropy as soon as the magnetization is driven slightly out of plane.
This may serve as an justification for the missing out-of-plane spin-torque component in the model of Slonczewski.

\section*{Conclusion}
The presented method for the solution of the Landau-Lifshitz-Gilbert equation including spin-torque effects is able to describe both the spin transport in multi-layer structures as well as current driven domain-wall motion.
The spin-torque in multi-layers as predicted by Slonczewski is essentially reproduced except for an out-of-plane component that is not present in the model by Slonczewski.
However, the out-of-plane component is shown to play an inferior role in typical multi-layer structures since it is usually almost completely compensated by shape anisotropy effects.
The presented method only requires the geometry and material parameters of the multilayer system as input.
In contrast the model of Slonczewski requires some global parameters that depend on the system as a whole and that have to be determined by experiment in order to perform meaningful simulations.
For the description of current-driven domain-wall motion the presented method is compared to a simplified model by Zhang and Li and shows a good agreement.
Small deviations can be explained by diffusion terms that are omitted in the simplified model.
While the present method is able to resolve the temporal evolution of the spin-accumulation exactly it can also be used for an adiabatic description since the time integrator for the spin-accumulation behaves very well even for large time steps.

\section*{Acknowledgements}
The financial support by
the Austrian Federal Ministry of Science, Research and Economy and the National Foundation for Research, Technology and Development
as well as
the Austrian Science Fund (FWF) under grant W1245,
the innovative projects initiative of Vienna University of Technology,
and the Royal Society under UF080837
is gratefully acknowledged.

\section*{Additional Information}
The authors declare no competing financial interests.


\section*{Contributions}
C.A. implemented the presented algorithm, performed the simulations, and wrote the main manuscript. M.R. and D.P. developed the algorithm and wrote the corresponding section. F.B., C.V., G.H., and D.S. developed the numerical experiments. G.H. and D.S. wrote the introduction. All authors reviewed the manuscript.
\end{document}